\documentclass[cameraready]{Interspeech}
\usepackage{microtype}
\usepackage{graphicx}
\usepackage{subfigure}
\usepackage{booktabs} 
\usepackage{amsmath}
\usepackage{adjustbox}
\usepackage{svg}
\usepackage[table]{xcolor}
\usepackage{algorithm}
\usepackage{algpseudocode}
\usepackage{hyperref}
\usepackage{amssymb}
\usepackage{mathtools}
\usepackage{amsthm}
\usepackage{xcolor}

\usepackage{amsmath}

\newcommand{\modelname}{TADA}

\title{\modelname: A Generative Framework for Speech Modeling via Text-Acoustic Dual Alignment}

\author[affiliation={1,2}, correspondingauthor]{Trung}{Dang}
\author[affiliation={1},correspondingauthor]{Sharath}{Rao}
\author[affiliation={1}]{Ananya}{Gupta}
\author[affiliation={1}]{Christopher}{Gagne}
\author[affiliation={1}]{Panagiotis}{Tzirakis}
\author[affiliation={1}]{Alice}{Baird}
\author[affiliation={1}]{Jakub Piotr}{Cłapa}
\author[affiliation={2}]{Peter}{Chin}
\author[affiliation={1}]{Alan}{Cowen}

\address{
    $^1$ Hume AI, USA \\
    $^2$ Dartmouth College, USA
}

\email{trungv.dang@outlook.com, sharath@hume.ai}


\usepackage{comment}

\begin{document}

\maketitle

\begin{abstract}
Modern Text-to-Speech (TTS) systems increasingly leverage Large Language Model (LLM) architectures to achieve scalable, high-fidelity, zero-shot generation. However, these systems typically rely on fixed-frame-rate acoustic tokenization, resulting in speech sequences that are significantly longer than, and asynchronous with their corresponding text. Beyond computational inefficiency, this sequence length disparity often triggers hallucinations in TTS and amplifies the modality gap in spoken language modeling (SLM). In this paper, we propose a novel tokenization scheme that establishes one-to-one synchronization between continuous acoustic features and text tokens, enabling unified, single-stream modeling within an LLM. We demonstrate that these synchronous tokens maintain high-fidelity audio reconstruction and can be effectively modeled in a latent space by a large language model with a flow matching head. Moreover, the ability to seamlessly toggle speech modality within the context enables text-only guidance--a technique that blends logits from text-only and text-speech modes to flexibly bridge the gap toward text-only LLM intelligence. Experimental results indicate that our approach achieves performance competitive with state-of-the-art TTS and SLM systems while virtually eliminating content hallucinations and preserving linguistic integrity, all at a significantly reduced inference cost.
\end{abstract}

\section{Introduction}
\label{introduction}

Recent advancements in Text-to-Speech (TTS) synthesis have been largely driven by the adoption of Large Language Model (LLM) architectures \cite{valle}. By leveraging the scaling laws observed in Natural Language Processing (NLP) \cite{kaplan2020scaling}, these systems have achieved unprecedented success in zero-shot voice cloning and high-fidelity speech generation. Central to these approaches is the discretization of speech into acoustic tokens, allowing the generation task to be framed as a next-token prediction problem within a unified transformer-based framework.

However, the efficacy of these models is fundamentally constrained by the structural discrepancy between text and audio representations. A majority of acoustic codecs \cite{encodec,dac,defossez2024moshi,higgsaudio2025} operate at a fixed frame rate to ensure signal fidelity. Because human speech contains far more acoustic information per second than linguistic information, the resulting speech sequences are often an order of magnitude longer than their corresponding text sequences. This frequency mismatch introduces a significant computational bottleneck; the disparate lengths of text and audio sequences drastically inflate the required context window for transformers, leading to a quadratic increase in computational complexity that reduces throughput during training and slows inference. Furthermore, the inherent lack of temporal alignment between low-frequency text tokens and high-frequency acoustic frames necessitates complex interleaving or hierarchical modeling strategies with semantic tokens. Such architectures prevent the realization of a truly single-stream model where linguistic and paralinguistic features are processed as synchronized facets of the same sequence. In interactive applications, this overhead elevates time-to-first-audio latency, which remains a primary barrier to natural, real-time human-AI interaction.

In this paper, we propose a novel tokenization schema that establishes a 1-to-1 synchronization between speech and text tokens. By utilizing an alignment-aware encoder-decoder architecture within a Variational Autoencoder (VAE) framework, we compress acoustic features into latent vectors that map 1-to-1 to discrete text units. This synchronized representation enables unified, single-stream modeling within an LLM, effectively treating text and speech as parallel tracks of a single information stream.

Our primary contributions are summarized as follows\footnote{Our code and pre-trained models are available at \url{https://github.com/HumeAI/tada}.}:
\begin{enumerate}
    \item Synchronous Tokenization: We introduce a method to extract acoustic features aligned one-to-one with text tokens, dramatically lowering the speech modeling frame rate. These tokens encapsulate full acoustic information, allowing for reconstruction independent of external conditions.
    \item Unified Autoregressive Modeling: We demonstrate that these synchronized features can be effectively modeled and generated in an autoregressive fashion, allowing for unified, single-stream modeling within an LLM to improve training throughput and inference efficiency.
    \item Modality Gap Mitigation: We propose Speech Free Guidance (SFG) to mitigate the modality gap caused by introducing audio to language models. By adjusting the logit scale between text-only and multimodal inference, we bridge this capability gap with minimal inference overhead.
\end{enumerate}

For convenience, we use ``acoustic features" and ``acoustic tokens" interchangeably, where ``tokens" encompasses both discrete and continuous forms. Our framework is designed as a unified speech-language model capable of concurrent text and speech generation. In this capacity, it can either function as a standalone TTS system or serve as a unified replacement for traditional, multi-stage LLM-TTS pipelines. Experiments demonstrate that our model performs on par with state-of-the-art TTS systems while achieving significantly higher speeds. Furthermore, the inductive bias from the 1:1 alignment between text tokens and speech features virtually eliminates content hallucinations. We also show that our model excels in speech-text co-modeling, with SFG bringing language performance close to that of text-only inference mode.

\section{Related Works}

\paragraph*{LLM-based TTS} The pivotal advancement enabling speech language model is the residual audio codec \cite{soundstream}, which discretizes audio into tokens that are amenable to autoregressive modeling. Whereas early models utilized a direct-to-acoustic-codes approach \cite{valle,livespeech,peng2024voicecraft}, more recent research introduce semantic tokens as an essential bridge to maintain global coherence \cite{borsos2023audiolm, soundstorm, speartts, zhang2023speechtokenizer, livespeech2, tortoise, cosyvoice, indextts2, xie2025fireredtts}.
By mutually conditioning content and style within an end-to-end system where a single LLM models both semantic and acoustic tokens \cite{defossez2024moshi, higgsaudio2025, peng2025vibevoice}, this framework improves model scalability and unlocks more diverse prosodic expressiveness across a wider range of contexts.

To enhance the efficiency of long-form synthesis and minimize the computational overhead of low-latency speech modeling, there has been a notable shift toward optimized temporal resolutions. This evolution is characterized by a transition from high-resolution multi-level discrete tokens utilizing rates such as 75 Hz \cite{livespeech}, 50 Hz \cite{zhang2023speechtokenizer,ye2025llasa}, or 25 Hz \cite{higgsaudio2025} to low-frame-rate codecs reaching 12.5 Hz \cite{defossez2024moshi, xie2025fireredtts,qwen3}. More recently, the field has explored continuous latent representations that achieve higher information density and further compress the sequence to 7.5 frames per second \cite{peng2025vibevoice}. By drastically reducing the number of LLM steps required for a second of audio, we can train these models at higher throughput, facilitate long-form generation without exceeding context limits, and enhance overall computational efficiency.

\paragraph*{Text-Speech Alignment in TTS} To ensure the reliability of production-grade systems, researchers have explored diverse alignment strategies. Non-LLM architectures often utilize hard alignment via explicit duration predictors (e.g., FastPitch \cite{fastpitch}) or soft alignment techniques like Monotonic Alignment Search (e.g., Glow-TTS \cite{kim2020glow}, VITS \cite{vits}). While these methods provide stability by modeling duration at the phoneme level, they often sacrifice the natural prosody and expressive flow characteristic of autoregressive models. In the context of LLM-based TTS, while recent works utilize semantic tokens as intermediate units to mitigate hallucinations and misalignments \cite{casanova2024xtts, peng2025vibevoice, xie2025fireredtts}, these representations do not inherently resolve the problem. The core challenge stems from a reliance on global attention mechanisms that lack an inductive monotonic bias. When coupled with the increasing prevalence of annotation errors in web-scale datasets--now reaching millions of hours--models frequently lose their temporal anchor, resulting in failures such as word skipping or content hallucinating. To mitigate this, LiveSpeech 2 \cite{livespeech2} introduces semantic guidance to enforce alignment during inference. Alternatively, some approaches adopt an interleaving strategy to fuse modalities at a fine-grained level during pre-training, as seen in SpiritLM \cite{nguyen2025spirit} and ELLA-V \cite{song2025ella}, while others introduce explicit monotonic constraints directly into the attention mechanism \cite{han2024vall} to stabilize the autoregressive process.

\paragraph*{Spoken Language Modeling} Several recent studies have moved toward modeling speech and text as a unified sequence, allowing for bidirectional conditioning between the two modalities. Representative works such as SpiritLM \cite{nguyen2025spirit} and LST \cite{lst} utilize interleaved pre-training to enable a single transformer backbone to natively process and generate both modalities. While these models represent a significant step toward cross-modal alignment, they often rely on low bit-rate discrete speech units (e.g., HuBERT semantic tokens \cite{hsu2021hubert}) as their primary vocabulary, which prioritize semantic or phonetic content over acoustic richness. Consequently, the modeled units often lack the resolution required to reconstruct high-fidelity, natural-sounding audio. Furthermore, models like LLaMA-Omni \cite{fang2024llama} attempt to bridge the gap by conditioning a speech decoder on the continuous hidden states of the LLM. However, because these synthesized features are generated in a separate branch and not recurrently integrated back into the autoregressive context, the system remains structurally decoupled. The LLM acts as a semantic director rather than an acoustic agent, lacking the granular feedback loop needed to perceive or adjust its own generated prosody and emotion. This architectural ``bottleneck" prevents the model from achieving a truly end-to-end realization where every acoustic detail is natively grounded in the LLM's internal state.

In contrast to the aforementioned LLM-based TTS and SLM approaches that often rely on explicit semantic tokens, fixed-rate speech tokens, and modality interleaving, we represent text and speech as synchronous dual components of a single stream. By adopting text tokens as the primary semantic unit, aligned with acoustic tokens spanning the full duration of each text token, our model enables autoregressive joint modeling with the efficiency of a text-only LLM, effectively bridging the representation gap between speech and text modalities in spoken language systems.

\section{Joint Speech-Text Tokenization}

Our tokenizer comprises three primary components: a temporal aligner, a token encoder, and an acoustic decoder. The aligner establishes a monotonic surjective mapping between LLM tokens and audio frames. The encoder then compresses the linguistic and contextual features into a single representative vector per token. Finally, the decoder utilizes these encoder outputs, augmented by the aligner's output positions, to reconstruct the high-fidelity audio signal. By utilizing LLM text tokens as the inherent semantic foundation, our framework bypasses the need for intermediary discrete semantic tokens, simplifying the generative pipeline.

\subsection{Aligner} 
\begin{figure}[h]
    \centering
    \includegraphics[width=\linewidth]{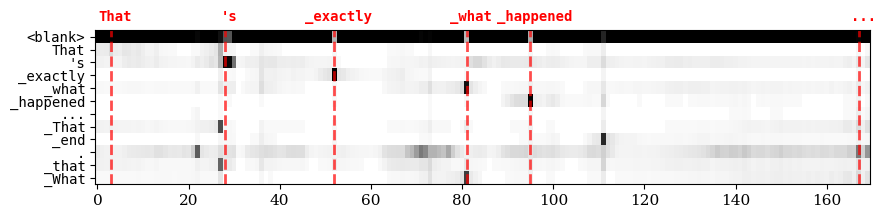}
    \caption{Word-level alignment via Viterbi decoding. An illustration of the forced alignment between a speech waveform and the transcript ``That's exactly what happened...". The Viterbi algorithm identifies the most probable frame-level assignment to determine a position $p_i$ for each token $w_i$ in the encoded text sequence.}
    \label{fig:alignment}
\end{figure}

The aligner processes a speech waveform and its corresponding sequence of text token IDs from the LLM vocabulary to generate a precise mapping between text tokens and audio frames. To achieve this, we train an acoustic model using Connectionist Temporal Classification (CTC) \cite{ctc} over the LLM’s subword vocabulary. The final alignment is extracted via forced alignment using the Viterbi algorithm, which identifies the most probable frame-level assignments for the target text. Figure \ref{fig:alignment} illustrates an example of this alignment process. Notably, this approach successfully aligns text tokens that do not appear in the top-k probability list (e.g., the token "That" in the provided example), ensuring the alignment remains robust to noisy audio and rare lexical tokens.

The aligner employs a transformer-based audio encoder based on Wav2Vec2 architecture \cite{wav2vec2}. Let $\boldsymbol{y^{\text{CTC}}}\in\mathbb{R}^{T\times V}$ denote the logits of the CTC backbone. We derive the audio-text alignment $p_i$, defined as the 1-based frame index corresponding to the text token $w_i$, as follows: $\boldsymbol{p}=\text{argmax}_{1\le p_1<\cdots<p_L \le T} \sum_{i=1}^L\boldsymbol{y}^{\text{CTC}}_{p_i,w_i}$, where $T$ is the acoustic frame length and $L$ is the text token length.

To effectively scale CTC training to large LLM vocabulary sizes, we introduce two complementary strategies. First, inspired by \cite{higuchi2022hierarchical}, we utilize an intermediate character-level CTC loss applied to internal hidden states to regularize the feature space and stabilize early convergence. Second, we implement a curriculum-based token selection mechanism, where the CTC loss is initially computed over a constrained subset of observed indices that dynamically expands as training progresses.

\begin{figure}[h]
    \centering
    \includegraphics[width=\linewidth]{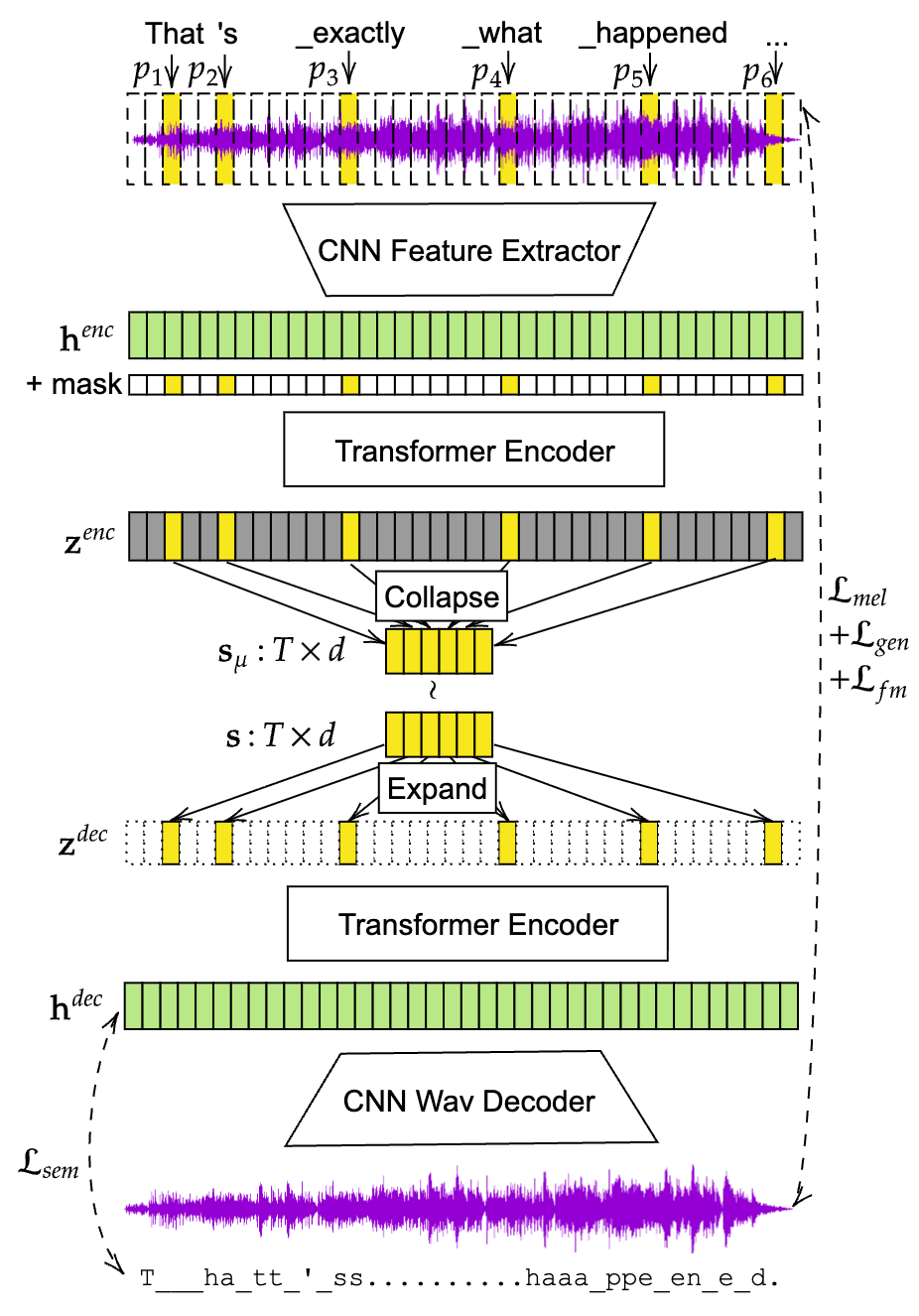}
    \caption{Operating under a Variational Autoencoder (VAE) framework, our model utilizes a symmetric encoder-decoder architecture. Each module integrates a CNN-based component for local acoustic feature extraction and reconstruction, complemented by a transformer-based backbone designed to capture the dynamic temporal range of synchronized speech-text sequences.}
    \label{fig:encdec}
\end{figure}

\subsection{Encoder} Figure \ref{fig:encdec} illustrates the encoder architecture. Unlike conventional fixed-rate encoders that typically rely on CNNs \cite{cnn} or RNNs \cite{lstm} to maintain rigid temporal alignment, our dynamic-rate approach employs a localized attention mechanism to aggregate variable-length audio segments into synchronized latent vectors. The encoder consists of two primary stages: a CNN-based feature extractor that projects raw audio into frame-level representations $\mathbf{h}^{\text{enc}} \in \mathbb{R}^{T \times d}$, and a transformer-based encoder that utilizes aligner outputs $\mathbf{p}$ to derive alignment-aware features $\mathbf{z}^{\text{enc}} \in \mathbb{R}^{T \times d}$. Within this transformer stage, the frame-level features are augmented with a binary indicator bit, where $1$ denotes a text-assigned position and $0$ represents a non-assigned position, which serves as a structural signal. This signal guides the multi-head attention mechanism to concentrate acoustic information specifically into the text-aligned positions, effectively anchoring the acoustic features to their corresponding text units.

We then condense the sequence of $T$ feature vectors by extracting only the text-aligned positions, resulting in a sequence of $L$ vectors $\mathbf{s}_\mu \in \mathbb{R}^{L \times d}$ that represent the latent means for each linguistic token. In this framework, $\mathbf{s}_\mu$ serves as the predicted mean of the latent distribution for each speech token, with the standard deviation fixed at $\sigma_0=0.5$. Following VibeVoice \cite{peng2025vibevoice}, to ensure sufficient variance for robust autoregressive modeling, we employ the reparameterization trick proposed in \cite{sun2024multimodal} to sample the latent representation $\mathbf{s} = \boldsymbol{s}_\mu + \boldsymbol{\sigma} \odot \boldsymbol{\epsilon}$, where $\boldsymbol{\epsilon} \sim \mathcal{N}(\mathbf{0}, \mathbf{I})$ and $\boldsymbol{\sigma}$ is sampled from $\mathcal{N}(0, k_{\sigma}\sigma_0)$, where $k_{\sigma}\ge 1$ is a constant. 

\begin{figure}[t]
    \centering
    \includegraphics[width=\linewidth]{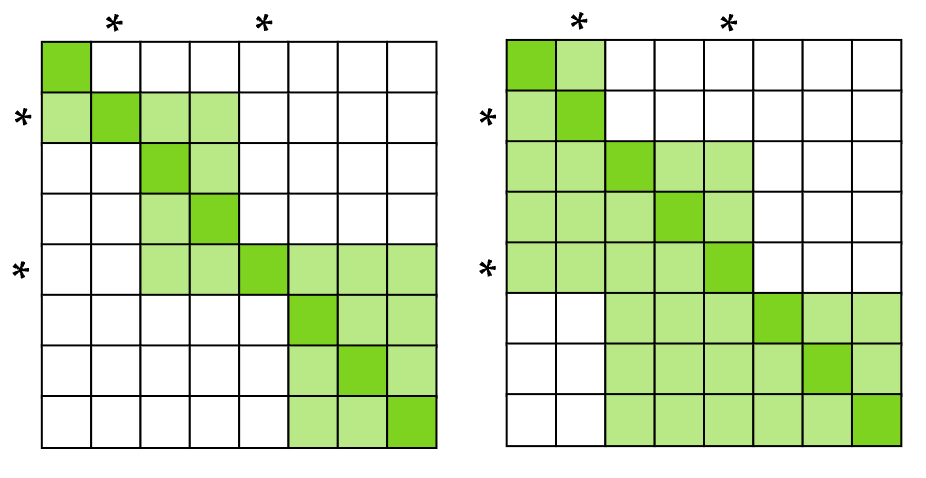}
    \caption{Attention mask of the encoder (left) and the decoder (right). Asterisks ($*$) mark text-assigned temporal indices. In the encoder, non-text-assigned positions are restricted to intra-block attention, excluding boundary tokens; conversely, text-assigned positions are permitted to attend across both preceding and succeeding blocks. The decoder also utilizes a localized mechanism where each position attends to the current and immediately preceding blocks.}
    \label{fig:attn}
\end{figure}

Additionally, to ensure a strictly local information flow, we implement an information bottleneck via a constrained attention mask: frames at a text-assigned position $p_i$ are permitted to attend only to frames within the temporal range $[p_{i-1} + 1, p_{i+1} - 1]$. Similarly, non-assigned frames $p$, where $p_i < p < p_{i+1}$, are restricted to the local window $[p_i + 1, p_{i+1} - 1]$. As illustrated in Figure \ref{fig:attn} (left), this masking strategy anchors acoustic information to its corresponding text unit, ensuring that the $i$-th token feature is derived exclusively from the audio segment within the indices $[p_{i-1}+1, p_{i+1}-1]$. By extracting features from these localized segments, the model produces temporally-grounded representations that facilitate efficient autoregressive modeling during speech generation.

\subsection{Decoder} We employ two decoders that are architecturally identical, differing only in their attention masks: the first is trained jointly with the encoder, while the second is trained with the encoder parameters frozen. Adopting a design complementary to the encoder, the decoder performs feature expansion on the tokens $\boldsymbol{s}\in\mathbb{R}^{L\times d}$, guided by positions $\boldsymbol{p}$, to produce a 50 Hz sparse sequence $\boldsymbol{z}^{\text{dec}} \in \mathbb{R}^{T \times d}$, defined such that $\mathbf{z}^{\text{dec}}_t = \mathbf{s}_i$ if $t = p_i$ and $\mathbf{z}^{\text{dec}}_t = \mathbf{0}$ otherwise. This sequence is transformed into a dense representation via a Transformer before a multi-layer CNN-based module synthesizes the final raw waveform. The first decoder utilizes global attention and is trained jointly with the encoder to ensure the tokens capture sufficient information for high-fidelity reconstruction. Subsequently, keeping the encoder and initial decoder parameters fixed, we train a second streamable decoder using a local attention schema. This configuration enables efficient autoregressive decoding by requiring only the KV-cache of the most recent segment, while computation within each individual segment remains non-autoregressive. The specific attention mask for this streaming mechanism is illustrated in Figure \ref{fig:attn} (right); here, each position attends to a range defined by $[p_{i-2}+1, p_i]$, where $p_i$ represents the first aligned position following the current frame (excluding itself).

\subsection{Training Loss} The VAE is optimized using a composite objective function comprising five distinct terms: a multi-scale mel-spectrogram loss, adversarial losses, a multi-scale feature matching loss, a semantic loss, and a Kullback–Leibler (KL) divergence term. The multi-scale mel-spectrogram loss ($\mathcal{L}_{\text{mel}}$) minimizes the $L_1$ distance between log-mel spectrograms calculated across varied window and hop sizes to capture both temporal transients and spectral envelopes. To achieve high-frequency realism, we employ adversarial losses, specifically a generator loss ($\mathcal{L}_{\text{gen}}$) and a discriminator loss ($\mathcal{L}_{\text{disc}}$), which drive the synthesized audio toward the manifold of natural speech. Complementing this, a multi-scale feature matching loss ($\mathcal{L}_{\text{fm}}$) computes the $L_1$ distance between the intermediate hidden feature maps of the discriminators for real versus reconstructed audio, providing a structural signal that stabilizes GAN training \cite{hifigan}. Furthermore, we introduce a semantic loss ($\mathcal{L}_{\text{sem}}$) by employing a linear projection head atop the decoder’s transformer backbone to predict the CTC-aligned grapheme sequence; this auxiliary objective ensures that the latent representations preserve strong linguistic grounding. The final training objective is defined as: $\mathcal{L}_{\text{total}} = \lambda_{\text{mel}}\mathcal{L}_{\text{mel}} + \lambda_{\text{gen}}\mathcal{L}_{\text{gen}} + \lambda_{\text{fm}}\mathcal{L}_{\text{fm}} + \lambda_{\text{sem}}\mathcal{L}_{\text{sem}} + \lambda_{\text{KL}}\mathcal{L}_{\text{KL}}$, where $\lambda_{*}$ denotes the scalar weights assigned to each component. The term $\mathcal{L}_{\text{KL}} = |\boldsymbol{\mu}|^2$ represents the KL divergence assuming a standard normal prior $\mathcal{N}(\mathbf{0}, \mathbf{I})$ and a fixed variance, with constant terms omitted for brevity.

\section{TADA: Text-Acoustic Dual-Alignment Modeling}

\begin{figure}
    \centering
    \includegraphics[width=0.5\textwidth]{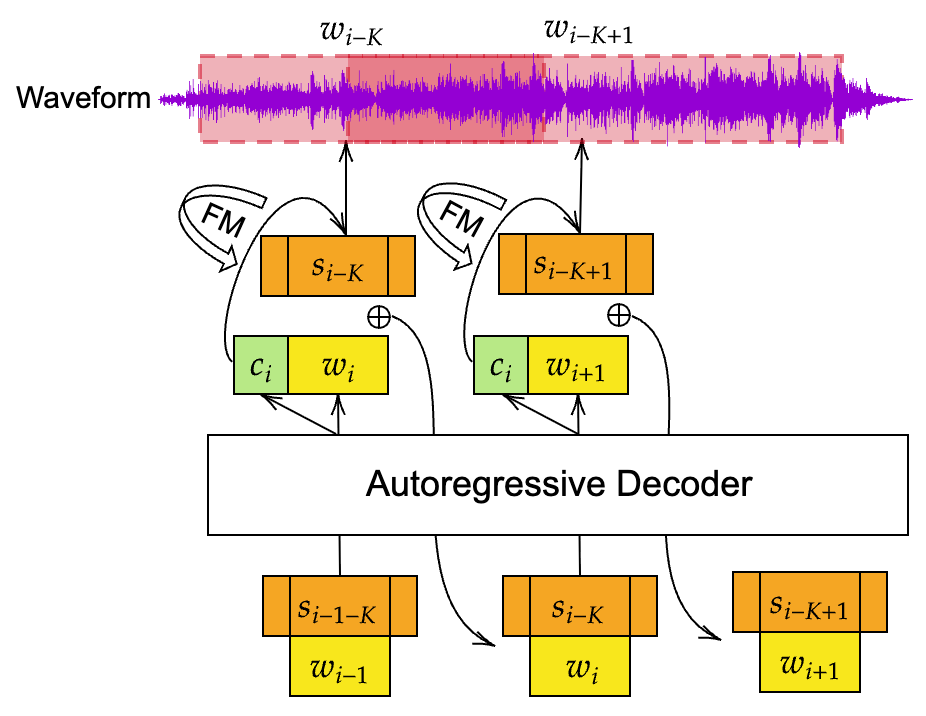}
    \caption{Each text token $w_i$ is paired with the speech representation at the $K$-shifted position, comprising token features $s_{i-K}$, the number of preceeding frames $f_{i-K}^{\text{before}}$, the number of successive frames $f_{i-K}^{\text{after}}$, and processed by an autoregressive decoder. The decoder predicts the next text token and produces a conditioning vector, which the flow matching head uses to generate the next speech representation $\left(s_{i-K+1},f_{i-K+1}^{\text{before}},f_{i-K+1}^{\text{before}}\right)$.}
    \label{fig:arch}
\end{figure}

In this section, we detail the architecture of \modelname{}, which primarily consists of a Large Language Model (LLM) integrated with a flow matching head. Figure \ref{fig:arch} illustrates our overall architecture.

\subsection{LLM Backbone} Our architecture integrates acoustic information by modifying a conventional LLM with additional acoustic embeddings. We perform additive fusion of text and acoustic input embeddings at each step; the resulting hidden states from the final layer are then projected into an acoustic output space to serve as the conditioning signal for the diffusion head, in addition to being used for text logits as usual. To allow for text lookahead, we shift the acoustic features by $K$ positions; that is, the text token at position $i$ is coupled with the acoustic features for the text token at position $i-K$. In contrast to prior works that interleave speech and text \cite{nguyen2025spirit,lst}, we combine them into a single stream. This approach maximizes the audio temporal context for a fixed sequence length, increasing training and inference throughput while preserving the native text-only inference paradigm. For instance, in an interleaved setup where text is sampled at 3 tokens per second and speech at 25 tokens per second, a 2048-token context window is exhausted in just 73 seconds of audio. In contrast, by synchronously modeling text and speech, our approach can accommodate 682 seconds of audio within the same token budget, a nearly ten-fold increase in information density. The last hidden state of each LLM step is fed into both a language modeling head to generate the text token distribution and a flow matching head to serve as the conditioning signal for generating the text-aligned acoustic representation.

\subsection{Flow Matching Head} The flow matching head jointly predicts the acoustic features and the temporal duration associated with each text token at step $i$, conditioning on the hidden representation $\boldsymbol{c}_i \in \mathbb{R}^{d_c}$ obtained from the LLM output. To model discrete frame durations, we employ Bit Diffusion \cite{bitdiffusion}, utilizing gray coding \cite{gray1953pulse} to minimize the impact of bit-flip errors. Specifically, the model predicts the acoustic embedding along with two sets of $b$ elements representing the number of preceding and successive blank frames relative to the token, referred to as $f^{\text{before}}_i$ and $f^{\text{after}}_i$, respectively. Note that $f^{\text{after}}_i=f^{\text{before}}_{i+1}$. These target analog bits are represented as continuous values in $\{-1, 1\}$ during the diffusion process; upon inference, these are quantized into discrete bits to reconstruct the integer frame counts.

Formally, the composite target vector $\mathbf{y}_i \in \mathbb{R}^{d_c + 2b}$ is defined as:

\[\mathbf{y}_i = [\boldsymbol{s}_i, \text{Analog}(\text{Gray}(f^{\text{before}}_i)), \text{Analog}(\text{Gray}(f^{\text{after}}_i))]\]

Conditioned on the LLM hidden state $\boldsymbol{c}_i \in \mathbb{R}^{d_c}$, the flow matching head learns a vector field $v_{\theta}(\mathbf{y}_t, t \mid \mathbf{c}_i)$ that transforms a Gaussian noise distribution $p_0$ to the target distribution $p_1$ centered at $\mathbf{y}_i$ \cite{lipman2022flow}. The objective is to minimize:

\[\mathcal{L}_{\text{flow}} = \frac{1}{L}\sum_{i=1}^L\mathbb{E}_{t, \mathbf{y}_0, \mathbf{y}_i} \left[ || v_{\theta}(\mathbf{y}_t, t | \mathbf{c}_i) - (\mathbf{y}_i - (1 - \sigma_{\min})\mathbf{y}_0) ||^2 \right]\]

where $\mathbf{y}_0\sim p_0$, $\mathbf{y}_i \sim q(\mathbf{y})$ (the data distribution), $\mathbf{y}_t = t\mathbf{y}_i + (1 - (1 - \sigma_{\min})t)\mathbf{y}_0$, and $\sigma_{\text{min}}$ is a small constant for numerical stability. At inference time, the predicted analog bits for duration are decoded as: $\hat{f}_i = \text{Gray}^{-1}(\mathbb{I}[\hat{\mathbf{b}} > 0])$, where $\mathbb{I}$ is the indicator function that quantizes the continuous output back into discrete bits.

We train the model with the flow matching loss $\mathcal{L}_{\text{flow}}$ and two optional losses: text cross-entropy loss $\mathcal{L}_{\text{CE}}$ and text knowledge distillation loss $\mathcal{L}_{\text{KD}}$. The overall loss is computed as $\mathcal{L} = \lambda_{\text{flow}} \mathcal{L}_{\text{flow}} + \lambda_{\text{CE}} \mathcal{L}_{\text{CE}} + \lambda_{\text{KD}} \mathcal{L}_{\text{KD}}$.

During inference, we utilize classifier-free guidance with a scale $\lambda_{\text{CFG}}$. For the negative condition, we can either use a zero vector or the output of a negative instance where all text tokens have been removed from the context; we refer to the latter as text-free guidance (TFG).

\subsection{Streamable Rejection Sampling} Since each LLM step generates acoustic information corresponding to a full text token, online rejection sampling can be employed at the step level to steer generation away from low-quality outputs. At each step, the system can verify whether the sampled acoustic features and durations accurately represent the target token, or detect anomalies such as unintended speaker switching or background music. This real-time rejection capability is particularly vital for large-scale systems; despite robust annotation, models trained on in-the-wild data frequently inherit artifacts from noisy training samples. In this work, we specifically propose using online rejection sampling to ensure \textit{speaker consistency}. We employ a lightweight speaker embedding head that operates on the generated acoustic features, rejecting any samples where the embedding deviates significantly from the average speaker embedding derived from the prompt. Specifically, we train a simple 3-layer MLP, with hidden dimensions of 768, 768, and 192, to predict a segment's ground-truth speaker embedding from its acoustic features. During inference, we pre-compute a reference speaker embedding by mean-pooling the projections across the prompt frames. To select the best output, we project the acoustic features of multiple flow-matching candidates into this embedding space and rank them by cosine similarity to the reference. The highest-scoring candidate is then selected, ensuring the generated frames preserve the target speaker's identity with negligible computational overhead.

\subsection{Speech Free Guidance (SFG)}

While incorporating both text and speech should theoretically enhance language modeling via paralinguistic cues, the inclusion of speech often degrades performance \cite{nguyen2025spirit,lst}, a phenomenon known as the modality gap. \modelname{} mitigates this by facilitating the blending of text-only and audio-augmented context, where the latter simply adds acoustic features to the base text embeddings, keeping the same context length. Specifically, the logits at step $i$ are computed as $\mathbf{z}_i = (1 - \lambda_{\text{SFG}}) \mathbf{z}_i^{(\text{text-only})} + \lambda_{\text{SFG}} \mathbf{z}_i^{(\text{text-speech})}$, where $\mathbf{z}_i^{(\text{text-speech})}$ and $\mathbf{z}_i^{(\text{text-only})}$ represent the logits with and without speech conditioning, respectively. To support text-only inference, we employ stochastic audio segment dropout during training. A dedicated acoustic mask embedding is added at each timestep to explicitly signal whether the model should condition on multimodal inputs or rely solely on text. By modulating the hyperparameter $\lambda_{\text{SFG}}$, we can steer the model's linguistic performance toward its text-only baseline while retaining the ability to leverage speech context where necessary.

\paragraph*{Comparison to prior speech-text modeling works} In the evolving landscape of unified speech-text modeling, frameworks such as LLaMA-Omni \cite{fang2024llama,fang2025llama}, Spirit-LM \cite{nguyen2025spirit}, and LST \cite{lst} represent significant efforts to bridge the modality gap within a single autoregressive architecture. However, these works share a critical limitation: a reliance on low-fidelity semantic units (e.g., discrete HuBERT tokens). By design, these units partially or entirely discard nuanced acoustic information, such as pitch, prosody, and speaker identity, thereby constraining the expressiveness of the synthesized audio. As a result, a representational bottleneck persists: the core LLM remains decoupled from the generated output because acoustic information is never fed back into the model's context, leaving the system blind to its own sampled speech. Architecturally, these models typically utilize an asynchronous format, interleaving modalities sequentially or asynchronously merging them into a serial stream based on fixed empirical ratios. In contrast, TADA leverages informationally rich, continuous acoustic features to ensure high-fidelity speech reconstruction while fusing text and speech into a single synchronous stream. By establishing a structural one-to-one mapping via our proposed dual alignment, TADA provides a robust inductive bias that inherently precludes the temporal misalignments typical of interleaved systems. This design significantly optimizes context efficiency and potentially enhances language modeling performance by maintaining a structural parity with text-only format.

\section{Experimental Setup}

\subsection{Datasets} The model is trained on a large speech corpus comprising the LibriLight corpus \cite{librilight}, an English proprietary dataset tailored for conversational speech, and a multilingual proprietary dataset in seven languages: Chinese, French, Italian, Japanese, Portuguese, Polish, and German. We segment the data using voice activity detection (VAD) into utterances of up to 30 seconds, totaling 270k hours of English data and 635k hours of non-English data. We transcribe each VAD segment using Parakeet-TDT-0.6b-v2 for English and European languages, and Whisper-V3 for Chinese and Japanese. Given the transcripts, we pre-extract alignment and token vectors prior to training. To mitigate potential transcription hallucinations, we filter out samples where a sequence of aligned positions spans more than three consecutive frames or where an aligned position gap exceeds 150 frames (3 seconds), as these metrics typically signal hallucinations, non-speech background, or missing text. This filtering can be performed dynamically during training using the alignment information.

\subsection{Tokenization} For the aligner, we employ Wav2Vec2-large \cite{wav2vec2} as the backbone, modifying the linear head to output 128,256 tokens of the Llama vocabulary instead of 32 graphemes. We train the aligner for 300k steps with a batch size of 64 and a sample duration of 30s. We utilize an exponentially decaying learning rate starting at $10^{-5}$, noting that a small learning rate is critical for convergence.

For the encoder, we employ a symmetric architecture consisting of a downsampling CNN backbone for feature extraction, followed by a 6-layer transformer. The decoder mirrors this design, utilizing a 6-layer transformer followed by an upsampling CNN backbone for raw audio synthesis. The transformer encoders employ a multi-head attention mechanism comprising 8 heads, a 1,024-dimensional hidden state, and a feed-forward network with a fourfold expansion factor, all integrated with Rotary Positional Embeddings (RoPE) \cite{rope}. The CNN components follow the Descript Audio Codec (DAC) \cite{dac} architecture. with strides of $(6, 5, 4, 4)$ to downsample every second of a 24kHz audio signal into a sequence $50$ frame features and upsample in the reversed order. Both the encoder and decoder transformers consist of 6 layers with 8 heads, a hidden dimension of 1,024, and a feed-forward dimension of 4,096. The embedding dimension for the acoustic vector is set to 512. We clamp the KL loss with the lower bound of $0.5$ to prevent posterior collapse and apply a dropout to the latent to foster a more robust feature space and promote informational redundancy within the bottleneck, as suggested in \cite{calm}.

During evaluation, we utilize an Euler solver to sample from the flow-matching objective over $N_{\text{FM}}=10$ steps. We apply classifier-free guidance (CFG) exclusively to the acoustic features $\boldsymbol{s}$ with a scale of $\lambda_{\text{CFG}}=1.8$; for the number of frames $f$, we bypass guidance as we observe that applying CFG to these discrete targets occasionally leads to sampling instability. If not specified, we use zero as the negative condition for CFG.

\subsection{Models} We adopt the Llama 3.2 1B and 3B base models \cite{llama3} for continued training on multimodal text-speech data. Specifically, we initialize the main Transformer decoder with Llama 3.2 weights and conduct training in two stages: (1) an initial phase, where the model is trained with a maximum context length of 192 for around 200k steps at a global batch size of 256; and (2) a second phase of around 200k steps, where the maximum context length is extended to 1,024 for the 1B variant and 2,048 for the 3B variant, while the global batch size is reduced to 64. Throughout both phases, we apply a cross-entropy loss (weight $\lambda_{\text{CE}}=0.05$) to adapt the model to the nuances of spoken text, alongside a KL divergence loss (weight $\lambda_{\text{KD}}=0.05$) against their base Llama model to preserve prior knowledge.

\subsection{Evaluation} We evaluate our models and baselines on a voice cloning benchmark using reference audio and target text pairs from following sources: (1) the SeedTTS-Eval test set \cite{seedtts}, comprising 1,088 samples from Common Voice; (2) a random subset of 1,002 samples from the LibriTTS-clean test set \cite{libritts} with durations exceeding 10s; and (3) a random subset of freeform speech samples from the Expressive Anechoic Recordings of Speech (EARS) dataset \cite{richter2024ears}, excluding those with long silence. For (1) and (2), we utilize the officially provided normalized transcripts, while for (3), we obtain the transcript via the Parakeet-TDT-0.6B model. For objective evaluation, we assess the Character Error Rate (CER) using the Parakeet-TDT-1.1B\footnote{While using Whisper-Large-v3 for CER is standard in related literature, we utilize Parakeet-TDT-1.1b to mitigate catastrophic hallucinations and achieve more stable evaluation scores.} model \cite{parakeet}, Speaker Similarity (SIM) via the VoxSim model \cite{voxsim}, and Objective Mean Opinion Score (oMOS) using UTMOSv2 \cite{utmosv2}. These metrics and models were selected because they are primarily optimized for correlation with human perception rather than relying solely on self-supervised objectives. To mitigate potential biases related to duration, temporal alignment, or leading and trailing silence, SIM is computed by comparing the first 5 seconds of the reference with the final 5 seconds of the output for reconstruction, and the middle 5 seconds of both for voice cloning. Similarly, we restrict oMOS evaluation to the center 5 seconds of each sample. For subjective evaluation, we ask human raters to score each sample on speaker similarity (sSIM) and naturalness (sMOS). We instruct raters to pay attention to prosody and intonation and to judge whether the tone matches the context of the spoken text. We obtain ratings for each metric on a 1--5 scale. To reduce individual rater bias, each sample is rated by five raters and we report the average score. 

For language capability evaluation, we benchmark our models and baselines across two primary axes. First, we measure perplexity on conversations in the Seamless Interaction (SI) dataset \cite{agrawal2025seamless}, a collection of dyadic, face-to-face interactions designed to model natural conversational dynamics and social behaviors. Because this dataset consists of natural, spontaneous dialogue, it allows us to quantify the model's foundational predictive accuracy and its grasp of long conversational flow. Second, we evaluate performance on Spoken StoryCloze (sSC) and Spoken TopicStoryCloze (tSC) \cite{hassid2023textually}, which task the model with identifying the correct ending or topic of a spoken narrative. These benchmarks are selected to explicitly assess the model's long-range contextual understanding, commonsense reasoning, and ability to maintain semantic coherence over extended spoken sequences. For the evaluation of our models, we utilize text transcribed by the Parakeet-0.6B-v2 model to better align with the textual distribution of the training data.

\section{Results}

\subsection{Token Reconstruction Quality}

\begin{table}[h]
\setlength{\tabcolsep}{3pt}
\caption{Comparison of reconstruction performance across discrete and continuous tokenizers. We report the frame rate (fps), bits per frame (bpf), and key performance metrics (CER, SIM, and oMOS). Continuous baselines ($\dagger$) are reported at 16-bit precision.}
\label{tab:reconstruction}
\vskip 0.15in
\begin{center}
\begin{small}
\begin{sc}
\begin{tabular}{lccccc}
\toprule
Models & FPS & BPF & CER$\downarrow$ & SIM$\uparrow$ & oMOS$\uparrow$ \\
\midrule
Reference & 24k & 384 & 0.00 & 85.4 & 3.17 \\
\midrule
Encodec \cite{encodec,siuzdak2023vocos} & 75 & 40 & 0.16 & 82.4 & 3.04 \\
Encodec \cite{encodec,siuzdak2023vocos} & 75 & 80 & 0.11 & 83.6 & 3.10 \\
DAC-24khz \cite{dac} & 75 & 80 & 0.17 & 83.4 & 2.94 \\
Mimi \cite{defossez2024moshi} & 12.5 & 352 & 0.14 & 83.9 & 3.04 \\
Higgs V2 \cite{higgsaudio2025} & 25 & 80 & \textbf{0.10} & 84.1 & 3.20 \\
VibeVoice$^\dagger$ \cite{peng2025vibevoice} & 7.5 & 3k & 0.15 & \textbf{84.7} & 3.12 \\
\midrule
\modelname-Codec$^\dagger$ & \textbf{2--3} & 8k & 0.14 & 83.6 & \textbf{3.34} \\
\bottomrule
\end{tabular}
\end{sc}
\end{small}
\end{center}
\vskip -0.1in
\end{table}

\begin{table*}[t]
\caption{Voice cloning evaluation. Objective comparison between \modelname{ } and baseline models using character error rate (CER), speaker similarity (SIM) and Objective Mean Opinion Score (oMOS) metrics.}
\label{tab:tts}
\vskip 0.15in
\begin{center}
\begin{small}
\begin{sc}
\begin{tabular}{lccccccccccc}
\toprule
& & & & \multicolumn{3}{c}{SeedTTS-Eval} & \multicolumn{3}{c}{LibriTTSR-Eval} \\
Models & Mem & RTF & Hours & CER & SIM & oMOS & CER & SIM & oMOS \\
\midrule
XTTS v2 \cite{casanova2024xtts} & 2.1 & 0.19 & 30K & 2.10 & 70.3 & 2.67 & 0.59 & 74.0 & 3.17 \\
Index-TTS2 \cite{indextts2} & 10.1 & 0.58 & 55K & 0.31 & 79.8 & 2.95 & 0.23 & 83.3 & 3.34 \\
Higgs Audio v2 \cite{higgsaudio2025} & 16.3 & 0.44 & 10M & 9.57 & 75.3 & 2.98 & 1.88 & 79.7 & 3.28 \\
VibeVoice 1.5B \cite{peng2025vibevoice} & 5.2 & 0.51 & 0.5--1.6M & 3.07 & 72.3 & 2.92 & 1.25 & 79.5 & 3.24 \\
FireRedTTS-2 \cite{xie2025fireredtts} & 12.2 & 0.76 & 1.1--1.4M & 0.81 & 75.1 & 2.96 & 1.44 & 81.2 & 3.28 \\
\midrule
\modelname{}-1B & 17.4 & 0.09 & 270K & 0.73 & 77.9 & 2.79 & 0.55 & 80.2 & 3.11 \\
\modelname{}-3B-ML & 26.1 & 0.13 & 900K & 0.76 & 75.1 & 2.85 & 0.40 & 79.9 & 3.17 \\
\bottomrule
\end{tabular}
\end{sc}
\end{small}
\end{center}
\vskip -0.1in
\end{table*}

We begin by evaluating the reconstruction quality of our tokenization schema to determine its effectiveness as a modeling target for generative tasks. Table \ref{tab:reconstruction} compares reconstruction quality between \modelname-Codec and various baselines. Despite the inherent challenge of handling dynamic frame expansion, \modelname-Codec performs on par with the fixed-rate tokenizers used in state-of-the-art TTS models, suggesting that these learned representations are well-suited for use as modeling targets.

\subsection{Text-to-Speech}

\paragraph*{Voice Cloning Results} The results in Table \ref{tab:tts} highlight the performance of our synchronous tokenization schema in the voice cloning setting on the SeedTTS-Eval and LibriTTSR-Eval. Unlike models that rely on high-frequency acoustic tokens and often suffer from stochastic hallucinations, the 1-to-1 mapping in \modelname{} ensures high-fidelity generation with high reliability. Our approach matches the stability of specialized two-stage models trained on small and cleaner data (e.g., XTTS-v2, Index-TTS2), evidenced by a low CER, while remaining on par with top-tier cloning systems in terms of speaker similarity (e.g., Higgs Audio v2, VibeVoice). We also measured the content hallucination rate by identifying samples with a Character Error Rate (CER) exceeding $0.15$--a threshold representing unintelligible speech, skipped text, or inserted content. Under these criteria, our model exhibited zero hallucinations. In contrast, the number of hallucinated samples was $41$ for FireRedTTS-2, $24$ for Higgs Audio V2, and $17$ for VibeVoice 1.5B. Although these failures occur in a relatively small subset of samples, such catastrophic outliers can significantly compromise the reliability of production-grade applications. Our lower oMOS scores given competitive CER and SIM suggest that further improvements in perceptual audio quality could be realized through a more robust decoder specifically designed to adapt to the LLM output distribution (e.g., a flow matching module that learns a generative flow into a high-quality speech distribution). 

\begin{table}[t]
\caption{Performance comparison for long-form, expressive speech generation using EARS dataset samples.}
\setlength{\tabcolsep}{2pt}
\label{tab:longform}
\vskip 0.15in
\begin{center}
\begin{small}
\begin{sc}
\begin{tabular}{lccccc}
\toprule
Models & CER & SIM & oMOS & sSIM & sMOS \\
\midrule
Index-TTS \cite{indextts2} & 1.90 & 76.9 & 2.84 & 4.25 & 3.61 \\
VibeVoice 1.5B \cite{peng2025vibevoice} & 2.51 & 73.3 & 2.54 & 3.92  & 3.91 \\
FireRedTTS-2 \cite{xie2025fireredtts} & 21.6 & 73.8 & 2.84 & 3.98 & 3.58 \\
\midrule
\modelname{}-3B & 2.34 & 67.0 & 2.86 & - & - \\
+ Text-Free Guidance & 4.30 & 72.4 & 2.87 & - & - \\
+ Online RS & 2.74 & 74.7 & 2.84 & 4.18 & 3.78 \\
\bottomrule
\end{tabular}
\end{sc}
\end{small}
\end{center}
\vskip -0.1in
\end{table}

Table \ref{tab:longform} presents our voice cloning results for long-form, expressive generation using the EARS dataset. In contrast to short-form generation, we observed occasional speaker drifting, which is reflected in the SIM scores. This issue is significantly mitigated by applying text-free guidance and online rejection sampling, resulting in a SIM score second only to IndexTTS. In subjective evaluations, our method ranks second in both speaker similarity and naturalness, indicating that our approach is more expressive than models trained on clean data, such as IndexTTS, while maintaining higher fidelity than models trained on fixed-rate tokens.

\paragraph*{Inference Time \& Memory Analysis} We measured the RTF and peak GPU memory (GB) of the baselines and our models by averaging metrics over inference on a single long target text using 49 different 5-second speech prompts, including the time and memory required for processing prompts with pre-extracted transcripts. All measurements were performed on a single H100 GPU with typical audio outputs of 20 seconds, as shown in Table \ref{tab:tts}. By leveraging a reduced frame rate of $2$--$3$ fps, \modelname{} achieves a significantly lower RTF than systems generating fixed discrete acoustic frames, despite the added overhead of text-acoustic alignment for the prompt. Furthermore, while VibeVoice \cite{peng2025vibevoice} also models continuous features at $7.5$ fps, our architecture circumvents costly step-wise conversions between semantic, acoustic, and audio features to achieve a substantial speedup. While \modelname{} exhibits a slightly higher peak memory footprint, partially due to the transcription model and wav2vec 2.0 Large-based aligner used for prompt processing, these processed prompts can be cached and reused across sessions. Furthermore, we expect memory consumption to scale more efficiently than competing models as output duration increases, owing to its more compact context length.

\subsection{Spoken Language Modeling}

\begin{table}[t]
\caption{Language capability evaluation on the Seamless Interaction (SI), sSC, and tSC datasets, comparing text-only (T) and text-speech (TS) inference modes with and without Text-Only Guidance.}
\setlength{\tabcolsep}{3pt}
\label{tab:speech-text-eval}
\vskip 0.15in
\begin{center}
\begin{small}
\begin{sc}
\begin{tabular}{lcccccc}
\toprule
 & \multicolumn{2}{c}{SI PPL$\downarrow$} & \multicolumn{2}{c}{sSC$\uparrow$} & \multicolumn{2}{c}{tSC$\uparrow$} \\
Models & $T$ & $TS$ & $T$ & $TS$ & $T$ & $TS$ \\
\midrule
Llama-1B-Ins & 23.7 & - & 73.8 & - & 97.4 & - \\
Llama-3B-Ins & 20.9 & - & 79.4 & - & 98.0 & - \\ \midrule
TWIST-7B \cite{hassid2023textually} & - & - & - & 55.3 & - & 74.1 \\
TWIST-13B \cite{hassid2023textually} & - & - & - & 55.4 & - & 76.4 \\
SpiritLM-7B \cite{nguyen2025spirit} & - & - & 79.4 & 61.0 & 98.0 & 82.9 \\
\midrule
\modelname{}-1B & 21.4 & 25.4 & 66.5 & 59.1 & 93.7 & 82.0\\
\modelname{}-3B-ML & 19.6 & 22.7 & 66.8 & 60.1 & 94.2 & 84.2 \\
\modelname{}-3B-ML w/ SFG & - & \textbf{20.5} & - & \textbf{66.7} & - & \textbf{94.7} \\
\bottomrule
\end{tabular}
\end{sc}
\end{small}
\end{center}
\vskip -0.1in
\end{table}

Table \ref{tab:speech-text-eval} compares our models against the base Llama model concerning perplexity on the Seamless Interaction dataset, and against two additional baselines, TWIST \cite{hassid2023textually} and Spirit-LM \cite{nguyen2025spirit}, on the sSC and tSC benchmarks. Notably, these baselines have a minimum size of 7B parameters. Each model is evaluated in text-only (T) or text-speech (TS) mode where applicable. Regarding perplexity, our model demonstrates strong performance in text-only mode, outperforming the base Llama models due to being fine-tuned on the spoken text distribution. When speech is modeled alongside text, perplexity increases slightly but remains within a reasonable range, with \modelname{}-3B-ML achieving better results than the Llama-1B-Instruction model. On sSC and tSC, we observe an accuracy drop when comparing \modelname{} operating in text-only mode to the base Llama models, suggesting the model trades off some of its text pretraining capacity to accommodate the speech modality. In text-speech mode, our model slightly trails Spirit-LM on the sSC dataset but outperforms it on tSC. This is particularly notable given our model's smaller size, fewer decoding steps, and continuous acoustic output, whereas SpiritLM-7B models \textit{semantic} tokens of small vocabulary size at 50Hz.

Crucially, applying text-only guidance with $\lambda_{\text{SFG}} = 0.5$ enables our model to surpass all other text-speech baselines, closely approaching text-only accuracy. Interestingly, on tSC, the accuracy of text-speech mode with text-only guidance slightly exceeds both the pure text-only and standard text-speech modes. This suggests that while the presence of speech can strain pure linguistic capacity, it may provide valuable contextual cues that aid in overall conversational understanding. During inference, SFG leverages parallel execution by injecting a text-only entry into the batch that exactly matches the speech-conditioned entry's context length and generation steps. While this doubles the effective batch size, the RTF overhead remains a negligible 0.01 for an original batch size of 1, ensuring high cost-efficiency.

\section{Analysis}

\begin{figure}[h]
    \centering
    \includegraphics[width=\linewidth]{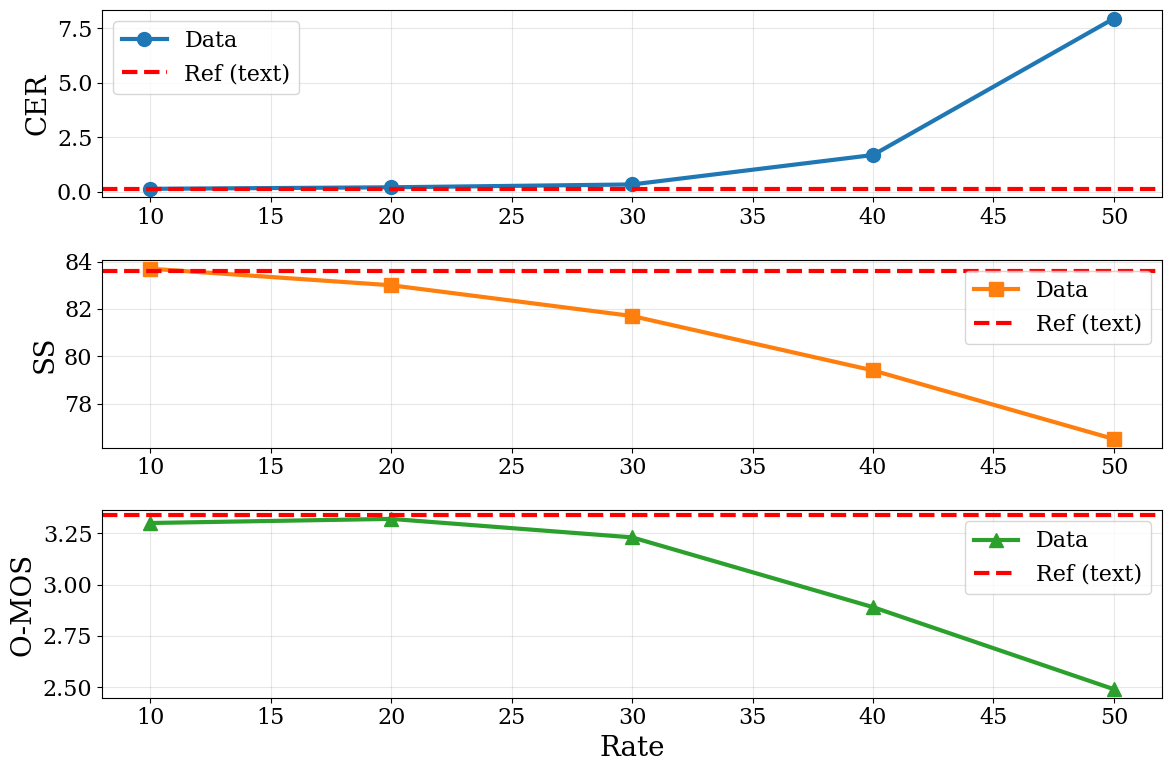}
    \caption{Reconstruction quality (CER, SS, and oMOS) vs. token density for uniformly spaced tokens.}
    \label{fig:ranges}
\end{figure}

\paragraph*{Fixed-Rate Tokenization} We test our token using uniformly spaced intervals rather than text-aligned positions. Figure \ref{fig:ranges} illustrates the scores for position intervals ranging from 10 to 50 frames (equivalent to 5 to 1 token per second). Notably, despite being trained exclusively on text-aligned positions, the decoder exhibits robustness in text-agnostic settings. This indicates that the acoustic features encode arbitrary audio segments rather than only those associated with text units. Our results demonstrate that high-fidelity reconstruction is maintained at rates as low as 2.5 tokens per second, suggesting these tokens are well-suited for fixed-rate encoding of non-transcribed, unaligned audio.

\paragraph*{Computational Overhead of the Diffusion Head}
We evaluate the computational overhead introduced by flow matching sampling at each LLM decoding step, with results summarized in Table \ref{tab:time}. Our analysis indicates that TTS quality converges between $4$ and $10$ steps, a range that increases the per-token latency compared to a text-only LLM step by approximately $50\%$ to $75\%$. Despite this non-trivial overhead, \modelname{} achieves a significant inference speedup because its frame rate is several times lower than the most frame rate efficient baselines.

\begin{table}[h]
\caption{Analysis of inference latency and TTS performance of \modelname{}-1B across varying flow matching sampling steps.}
\setlength{\tabcolsep}{3pt}
\label{tab:time}
\vskip 0.15in
\begin{center}
\begin{small}
\begin{sc}
\begin{tabular}{lcccccc}
\toprule
$N_{\text{fl}}$ & RTF & $T^{\text{step}}_{\text{LLM}}$ & $T^{\text{step}}_{\text{FM}}$ & CER & SIM & oMOS \\
\midrule
2 steps & 0.05 & 31 & 8 & 1.82 & 77.3 & 2.82 \\
4 steps & 0.05 & 31 & 14 & 0.85 & 79.1 & 2.96\\
10 steps & 0.09 & 31 & 23 & \textbf{0.55} & \textbf{80.2} & \textbf{3.11} \\
20 steps & 0.13 & 31 & 54 & 0.63 & 79.8 & \textbf{3.11} \\
\bottomrule
\end{tabular}
\end{sc}
\end{small}
\end{center}
\vskip -0.1in
\end{table}

\paragraph*{Language Preservation Losses} To analyze the trade-off between language preservation and audio quality, we conduct an ablation study focusing on two preservation losses: cross-entropy ($\mathcal{L}_{\text{CE}}$) and knowledge distillation ($\mathcal{L}_{\text{KD}}$). We trained three models for 100k steps: one with $\lambda_{\text{CE}}=0.1$, one with a higher weight of $\lambda_{\text{CE}}=1.0$, and one utilizing both losses at $\lambda_{\text{CE}}=\lambda_{\text{KD}}=0.1$, as in our base configuration. Our results indicate that while TTS performance remains consistent across all variants, perplexity increases with higher loss weights; notably, the best language preservation is achieved through the inclusion of the knowledge distillation loss.

\begin{table}[h]
\caption{Abalation results of \modelname-1B with different language preservation loss config.}
\setlength{\tabcolsep}{3pt}
\label{tab:speech-text-corpus}
\vskip 0.15in
\begin{center}
\begin{small}
\begin{sc}
\begin{tabular}{lcccccc}
\toprule
Models & PPL & CER & SIM & oMOS \\
\midrule
Llama-1B-Instruct & 23.7 & - & - & - \\
\midrule
\modelname{}-1B ($\lambda_{\text{CE}}=0.1$, Base) & 36.5 & 0.87 & 78.4 & 3.02 \\
\modelname{}-1B ($\lambda_{\text{CE}}=1.0$) & 31.2 & 0.89 & 77.6 & 3.00 \\
\modelname{}-1B ($\lambda_{\text{CE}}=\lambda_{\text{KD}}=0.1$) & 26.6 & 0.84 & 77.7 & 2.98 \\
\bottomrule
\end{tabular}
\end{sc}
\end{small}
\end{center}
\vskip -0.1in
\end{table}

\newpage

\section{Acknowledgement}

We would like to thank Taiga Ishida, Rashish Tandon, Georg Streich, Jeffrey Brooks for their insightful feedback and discussions throughout the development of the work. We are grateful to Hume AI for providing the high-performance computing resources to conduct our experiments.

\section{Generative AI Use Disclosure}

Generative AI tools are used for editing and polishing manuscripts, but not for producing a significant part of the manuscript.

\bibliographystyle{IEEEtran}
\bibliography{mybib}

\end{document}